\pdfoutput=1
\documentclass[conference]{IEEEtran}

\usepackage{ifthen}  
\usepackage{color} 

\usepackage{xcolor}  

\usepackage{amsmath}
\usepackage{amssymb}
\usepackage{amsthm}
\usepackage{mathpartir}

\usepackage{graphicx}
\graphicspath{{./figures/}}
\DeclareGraphicsExtensions{.pdf,.png}

\usepackage{listings}  

\usepackage{paralist}

\usepackage{cite} 
\usepackage{hyperref} 
\usepackage{cleveref}

\newcommand{\code}[1]{{\small\texttt{#1}}}

\newcommand{\term}[1]{\textbf{#1}}

\newcommand{\myparagraph}[1]{\smallskip \noindent{\textbf{{#1}.}}}
\newcommand{\myparagraphnodot}[1]{\smallskip \noindent{\textbf{{#1}}}}

\newcommand{\secref}[1]{$\S$\ref{sec:#1}}
\newcommand{\subsecref}[1]{$\S$\ref{subsec:#1}}

\newcommand{\appref}[1]{Appendix~\ref{app:#1}}

\newcommand{\figref}[1]{Figure~\ref{fig:#1}}
\newcommand{\tabref}[1]{Figure~\ref{tab:#1}}

\newcommand{\defref}[1]{Def~\ref{def:#1}}

\newcommand{\program}{\mathit{Prog}}
\newcommand{\archstate}{\mathit{Data}}
\newcommand{\microstate}{\mathit{Ctx}}
\newcommand{\htrace}{\mathit{HTrace}}
\newcommand{\ctrace}{\mathit{CTrace}}
\newcommand{\attack}{\mathit{Measure}}
\newcommand{\contract}{\mathit{Contract}}
\newcommand{\ctseq}{CT-SEQ}
\newcommand{\ctcond}{CT-COND}

\theoremstyle{definition}
\newtheorem{definition}{Definition}
\newtheorem{example}{Example}

\renewcommand{\program}{p}
\renewcommand{\archstate}{i}
\renewcommand{\microstate}{\mu}

\makeatletter
\newcommand{\xRightarrow}[2][]{\ext@arrow 0359\Rightarrowfill@{#1}{#2}}
\makeatother

\mathchardef\hyphenmathcode=\mathcode`\-

\lstdefinestyle{embedded}{
    basicstyle=\ttfamily\small,
    numbers=left,
    numberstyle=\tiny,
    numbersep=3pt,                  
    numberblanklines=true,
    frame=tb,
    aboveskip=5pt,
    belowskip=5pt,
    columns=fullflexible,
    showstringspaces=false,
    keepspaces=true,
    showlines=true,
    xleftmargin=8pt,
    backgroundcolor=\color{gray!4},
    framesep=1pt,
}
\lstdefinestyle{asm}{
    language={[x86masm]Assembler},
    basicstyle=\ttfamily\small,
    numbers=left,
    numberstyle=\tiny,
    numbersep=3pt,                  
    numberblanklines=true,
    frame=tb,
    aboveskip=0pt,
    belowskip=0pt,
    columns=fullflexible,
    showstringspaces=false,
    keepspaces=true,
    showlines=true,
    xleftmargin=8pt,
    backgroundcolor=\color{gray!4},
    morekeywords={MFENCE,R14,RAX,RBX,RCX,RDX,CMOVNL,CMOVNZ},
    morecomment=[l]{\#},
    commentstyle=\color{gray},
    keywordstyle=\ttfamily,
}

\lstdefinelanguage{yaml}{
    morecomment=[l]{\#},
    commentstyle=\color{gray}
}

\let\origlstlisting=\lstlisting
\let\endoriglstlisting=\endlstlisting
\renewenvironment{lstlisting}
{\mathcode`\-=\hyphenmathcode
\everymath{}\mathsurround=0pt\origlstlisting}
{\endoriglstlisting}

\newboolean{publicversion}
\setboolean{publicversion}{false}

\ifthenelse{\boolean{publicversion}}{
\newcommand{\grumbler}[2]{}
\newcommand{\editing}[1]{}

}{
\newcommand{\grumbler}[2]{\textcolor{red}{{{\bf #1}: #2}}}

\newcommand{\editing}[1]{{\color{red}#1}}

}

\newcommand{\status}[3]{}

\def\BibTeX{{\rm B\kern-.05em{\sc i\kern-.025em b}\kern-.08em
    T\kern-.1667em\lower.7ex\hbox{E}\kern-.125emX}}
\begin{document}
    
\title{Hide and Seek with Spectres: Efficient discovery of speculative information leaks with random testing} 

\author{\IEEEauthorblockN{Oleksii Oleksenko}
\IEEEauthorblockA{\textit{Microsoft Research}}
\and
\IEEEauthorblockN{Marco Guarnieri}
\IEEEauthorblockA{\textit{IMDEA Software Institute}}
\and
\IEEEauthorblockN{Boris Köpf}
\IEEEauthorblockA{\textit{Microsoft Research}}
\and
\IEEEauthorblockN{Mark Silberstein}
\IEEEauthorblockA{\textit{Technion}}
}

\maketitle

    \begin{abstract}

    Attacks like Spectre abuse speculative execution, one of the key performance optimizations of modern CPUs.
    Recently, several testing tools have emerged to automatically detect speculative leaks in commercial (black-box) CPUs.
    However, the testing process is still slow, which has hindered in-depth testing campaigns, and so far prevented the discovery of new classes of leakage.

    In this paper, we identify the root causes of the performance limitations in existing approaches, and propose techniques to overcome these limitations.
    With these techniques, we improve the testing speed over the state-of-the-art by up to two orders of magnitude.

    These improvements enable us to run a testing campaign of unprecedented depth on Intel and AMD CPUs. As a highlight, we discover two types of previously unknown speculative leaks (affecting string comparison and division) that have escaped previous manual and automatic analyses.

\end{abstract}

\begin{IEEEkeywords}
    Spectre, side-channel-attack, speculative-execution, random testing, constrained random verification
\end{IEEEkeywords}

    \section{Introduction}
\label{sec:intro}

Attacks like Spectre~\cite{Kocher2018} exploit speculative information leaks, which are a side-effect of performance optimizations in modern CPUs.
These attacks can ``trick'' a CPU into leaving traces of secret information that make it accessible to an adversary, bypassing program-level security checks.

So far, most of the speculative leaks 
(e.g., those underlying Spectre~\cite{Kocher2018}, Meltdown~\cite{Lipp2018}, MDS~\cite{RIDL,ZombieLoad}) have been discovered in a manual effort, by analysing public documentation, patents, and by laborious experimentation.

Recently, several tools have emerged to {\em automate} this slow and costly process:
{\em White-box} approaches~\cite{Trippel2018,Ghaniyoun2021,Fadiheh2019} analyse the processor specification and detect leaks early in the design process. They have so far been successfully applied to smaller open-source CPUs.
In contrast, {\em black-box} approaches~\cite{ Moghimi2020a,Xiao2020,Nemati2020,oleksenko2022revizor} analyze fabricated chips. They have already been applied to full-scale x86 CPUs, and are the focus of this paper.

In the absence of a specification, black-box approaches rely on two kinds of random testing, with different scope:
\begin{asparaenum}[(i)]
    \item {\em Template-based} tools~\cite{Moghimi2020a,Xiao2020} {\em mutate} code-templates known to trigger speculation. They can detect variants of known leaks rather than finding new ones.\looseness=-1
    \item {\em Model-based relational} tools~\cite{oleksenko2022revizor,Nemati2020} generate \emph{fully random} test cases (instruction sequences and inputs), and compare the leakage observed on the CPU with that specified by the leakage model (typically: an ISA-level specification of the permitted leaks).
\end{asparaenum}

Black-box approaches have successfully detected all known classes of speculative leaks, as well as previously unknown variants. However, so far they have {\em not discovered} any fundamentally new kind of speculative leakage. For template-based tools, this is explained by the scope of the approach. For model-based tools, this is due to the performance limitations of the existing approaches, which have been preventing in-depth testing---until now.

\myparagraph{Our approach}
In this paper, we present a model-based approach for detecting speculative leaks in black-box CPUs, which overcomes the performance limitations of state-of-the-art tools. We improve the
speed of testing by two orders of magnitude, which enables us to significantly increase the breadth and depth of our testing campaigns. In our experiments on Intel and AMD CPUs, we detected two
new classes of leaks that have escaped prior manual and automatic analysis.

Our key observation is that the vast majority of randomly-generated test cases are {\em ineffective} in that they have no chance of surfacing a speculative leak---yet existing tools~\cite{oleksenko2022revizor,Nemati2020} still include them in the expensive leakage analysis, e.g., based on symbolic execution.

This observation motivated us to (i) identify the characteristics of effective test cases; (ii) enforce generation of such test cases where possible; and (iii) design lightweight methods for filtering ineffective test cases. Thus, we are able to dramatically speed up the testing campaigns by either generating test cases that are effective by design, or pruning ineffective test cases {\em before} they undergo any expensive leakage analysis.

\myparagraph{Characteristics of effective test cases}
Our goal is to identify programs that speculatively leak more information than what the CPU leakage model prescribes.
Technically, this requires executing a program with {\em two} different inputs so that both executions {\em agree} on the leakage according to the leakage model (which we call {\em model trace}), but {\em differ} on the measurements made on the actual CPU (which we call {\em hardware trace}).
The following conditions are necessary for this to happen:
\begin{compactenum}
    \item The program misspeculates, i.e., there are {\em transient} instructions that are issued but never retire;\label{it:specfilter}
    \item The transient instructions affect the hardware trace, i.e., transient leakage becomes observable;\label{it:obsfilter}
    \item The program inputs result in identical model traces, which makes it possible to compare hardware traces that witness transient leakage.\label{it:booster}
\end{compactenum}

Our preliminary measurements show the potential of leveraging these conditions: with randomly generated programs and inputs, condition~\ref{it:specfilter} is satisfied in only 13\% of the cases, condition~\ref{it:obsfilter} is satisfied in only 7\% of the cases, and less than 6\% of the inputs satisfy condition~\ref{it:booster}.
It means that, overall, less than 0.5\% of program-input combinations are effective.

\myparagraph{Improving test case effectiveness}
We next explain how we check conditions~\ref{it:specfilter} and~\ref{it:obsfilter}, and how we ensure condition~\ref{it:booster} by design, with the following techniques:
\begin{asparaenum}
    \item \emph{Speculation filtering}: To check for misspeculation (condition~\ref{it:specfilter}) during execution of a test case, we leverage signals from the CPU's internal state by monitoring
	    speculation-related performance counters. This technique is inspired by~\cite{ragab2021rage} where it was used for manual analysis; here we leverage it for automated test case pruning.
    \item \emph{Observation filtering}: To check for visibility in the hardware trace (condition~\ref{it:obsfilter}), we modify the program by inserting a serialization fence after every instruction,
	    and then compare the execution of the original program with the serialized version. A discrepancy between the hardware traces indicates that the original test case encountered speculation that is visible in the trace.
    \item \emph{Contract-driven input generation}: To generate input pairs with identical model traces (condition~\ref{it:booster}), existing tools rely on accidental matches during random
	    sampling~\cite{oleksenko2022revizor} or on symbolic execution~\cite{Nemati2020}, which are wasteful and slow, respectively.  We propose a technique based on dependency tracking to
	    identify registers and memory locations that can be varied while keeping the leakage model traces identical. Compared to prior work, this technique turns out to be effective {\em and} fast.
\end{asparaenum}

\myparagraph{Implementation and experiments}
We implement these techniques on top a model-based tool called Revizor~\cite{oleksenko2022revizor}. Our implementation targets x86-64 CPUs.

To showcase the techniques, we run a testing campaign with 130 million test case executions, over 13 subsets of the x86 ISA.
The highlights are:
\begin{asparaitem}
    \item The techniques speed up testing by \emph{two orders of magnitude}. With the current state-of-the-art~\cite{oleksenko2022revizor}, this campaign would have required over \emph{2 months}. With the techniques we describe in this paper, the campaign took only \emph{16~hours}.
    \item During this campaign we discovered {\em two new speculative leaks} based on undocumented kinds of speculation in 64-bit division and string comparison operations. We describe them in Section~\ref{sec:findings}.
\end{asparaitem}

We further perform an in-depth analysis of how our techniques contribute to this result:
\begin{asparaitem}
    \item Speculation and observation filters consistently improve the testing speed. The largest speedups of an order of magnitude are achieved for programs generated from ISA subsets that {\em never} experience speculation, because virtually all test cases are filtered out. The smallest speed-ups are found in the instruction sets where speculation is ubiquitous (e.g., conditional branches that trigger Spectre V1), where a large portion of the test cases passes the filters.
    \item Contract-driven input generation reliably increases the number of detected leaks compared to random testing. This is because it increases the number of inputs satisfying condition
	    \ref{it:booster} to over
	    99\%, and achieves that with about 100x less performance overhead compared to symbolic execution-based input generation.
    \item For leaks that are already detected by the random baseline, our techniques {\em reduce the detection time by a factor of 10--50}.
    For the new speculative leaks, our techniques were instrumental for detecting them during our testing campaign.
\end{asparaitem}

The source code is publicly available under:

\begin{center}
    \url{https://github.com/microsoft/sca-fuzzer}
\end{center} 

\myparagraph{Responsible disclosure}
We disclosed our findings to Intel and AMD.
Both vendors acknowledged and confirmed the findings.

    \begin{figure}[tb]
    \center    \footnotesize
    \setlength{\tabcolsep}{2.7pt}
    \begin{tabular}{l|ccc|c}
                                                         & \multicolumn{4}{c}{Information Leakage Detected?}                                              \\
        \cline{2-5}
        x86-64 Subset                                    &                                                   & Intel        &              & AMD          \\
                                                         & Core6                                             & Core8        & Xeon         & Epyc         \\
        \hline
        \hline
        \code{cond}: Conditional branches                & \checkmark{}                                      & \checkmark{} & \checkmark{} & \checkmark{} \\
        \code{strn}: String operations                   & \checkmark{}                                      & \checkmark{} & \checkmark{} & \checkmark{} \\
        \code{dmul}: Division and multiplication         & \checkmark{}                                      & \checkmark{} & \checkmark{} & $\times$     \\
        \code{flag}: Operations on flags                 & $\times$                                          & $\times$     & $\times$     & $\times$     \\
        \code{lock}: Atomics w/ LOCK prefix              & $\times$                                          & $\times$     & $\times$     & $\times$     \\
        \code{atom}: Atomics w/o LOCK prefix             & $\times$                                          & $\times$     & $\times$     & $\times$     \\
        \code{dxfr}: Data transfer (load/store)          & $\times$                                          & $\times$     & $\times$     & $\times$     \\
        \code{setc}: Conditional byte set                & $\times$                                          & $\times$     & $\times$     & $\times$     \\
        \code{nop }: NOP instructions                    & $\times$                                          & $\times$     & $\times$     & $\times$     \\
        \code{logi}: Logical operations                  & $\times$                                          & $\times$     & $\times$     & $\times$     \\
        \code{conv}: Data type conversion                & $\times$                                          & $\times$     & $\times$     & $\times$     \\
        \code{cmov}: Conditional moves                   & $\times$                                          & $\times$     & $\times$     & $\times$     \\
        \code{bit }: Bit test and bit scan               & $\times$                                          & $\times$     & $\times$     & $\times$     \\
        \hline
        All except \code{cond}, \code{dmul}, \code{strn} & $\times$                                          & $\times$     & $\times$     & $\times$     \\
        \hline
    \end{tabular}
    \caption{Summary of the main testing campaign.
        The testing targets are:
        Intel i5-6500 (Core6);
        Intel i7-8650U (Core8);
        Intel E-2288G (Xeon);
        AMD EPYC 7543P (Epyc).
        All microcode patches were enabled.
        The campaign analyzed 13 subsets of the x86-64 ISA (details
	in~\appref{subsets}). Each subset was tested with 100'000 randomly
	generated programs, and each program executed with 100 inputs, amounting
	to 130 million test case executions per target.}
    \label{tab:results1}
\end{figure}

\section{Discovered Speculative Leaks}
\label{sec:findings}

In this section, we describe two new speculative leaks that we discovered with the techniques presented in this paper.
The discoveries became possible due to the improved testing speed, which enabled us to dramatically increase the depth and breadth with which we scrutinize the ISA, and made feasible the testing campaign in~\tabref{results1}.

This section describes only the leaks themselves; a reader interested in the \emph{process} of their discovery can find a detailed walk-through in~\appref{sco}.

\begin{figure*}[]
    \centering
    \includegraphics[width=.78\textwidth]{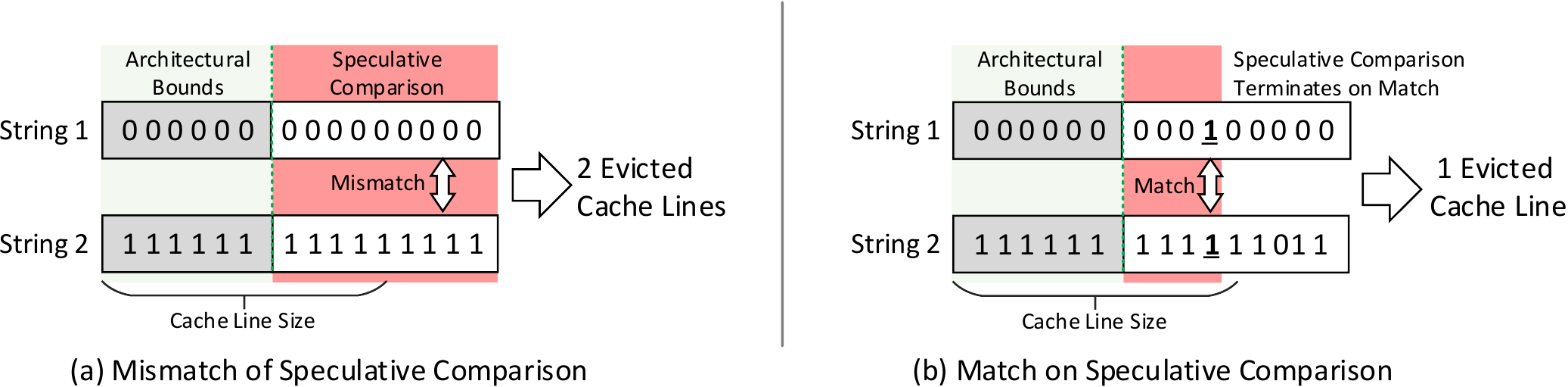}
    \vspace{-5pt}
    \caption{Illustration of leakage with String Comparison Overrun.}
    \label{fig:sco}
\end{figure*}

\subsection{SCO: String Comparison Overrun}

The first new speculative leak we found affects all four CPUs we tested (both Intel and AMD).
It is caused by string instructions prefixed by \code{REPE} (repeat while equal) and \code{REPNE} (repeat while not equal)\footnote{\code{REP} also speculates, but we found no information leaks caused by it.}.
With these prefixes, the following instruction is repeated the number of times indicated in the count register, or until the termination condition is met.

Our testing campaign revealed that the CPUs can speculatively compare (\code{CMPS}) or scan strings (\code{SCAS}) beyond the termination condition and leak the outcome into the cache state.
This new leak, which we call {\em String Comparison Overrun} (SCO), has two notable features:
\begin{compactenum}[(i)]
    \item It does not require training: The CPU always goes beyond the bounds if the counter value is not available;
    \item It does not require a software leakage gadget: A single instruction contains both the speculation trigger and the leaking memory access.
\end{compactenum}

For illustration, consider the following snippet, which compares two strings word-by-word for inequality, starting from the addresses in \code{rdi} (String 1) and \code{rsi} (String 2). The comparison terminates when it either finds a matching word or  reaches the bounds specified in the count register \code{rcx}.\looseness=-1

\begin{lstlisting}[style=embedded]
rcx = slow_computation()
REPNE CMPSW 
\end{lstlisting}

In the example, the value of \code{rcx} is not available, which triggers speculation, and the CPU compares memory beyond the specified bounds. Assuming the adversary controls String 1, it can perform equality checks on the memory following String 2 (see~\figref{sco}). The outcome affects the number of evicted cache lines: If the strings disagree on all words (\figref{sco}a), comparison continues until the speculation window expires, and if the window is large enough, at least two cache lines will be evicted. If not (\figref{sco}b), comparison terminates with the earliest matching word, and if the match is within the first cache line, only one line gets evicted. The attacker can distinguish these two cases via a cache side-channel attack, thus learning the outcome of out-of-bounds comparison.

This leak enables an adversary to 
determine the memory content following String 2 as far as speculation lasts. 
The complexity of determining memory content depends on the granularity: Word-level checks (\code{CMPSW}) require only up to $2^{16}$ attempts to produce a match whereas quad-word checks (\code{CMPSQ}) require up to $2^{64}$ attempts. The lower complexity, however, comes with smaller reach, as the number of comparisons is bounded by the speculation window: In our prototype implementation, we managed to leak 22 out-of-bounds bytes with \code{CMPSW} compared to 88 bytes with \code{CMPSQ}. 

\subsection{ZDI: Zero Dividend Injection}

The second speculative leak was discovered only in Intel CPUs, and it is triggered by division instructions.
Specifically, we found that 64-bit division can speculate on the value of one of its source operands.
We call it {\em Zero Dividend Injection} (ZDI), and to the best of our knowledge, this is the first documented case of value prediction in a commercial CPU.

For illustration, consider the following snippet, where \code{DIV} in line 2 divides a 128-bit operand in \code{rdx:rax} by a 64-bit operand in \code{rcx}.

\begin{lstlisting}[style=embedded]
rdx = slow_computation()
DIV rcx  // means rdx:rax/ rcx
MOV rsi, [array_base + rax]
// or MOV rsi, [array_base + rdx]
\end{lstlisting}

If values of registers \code{rcx} and \code{rax} are available but that of \code{rdx} is not, the CPU bypasses this dependency by speculatively assuming that \code{rdx} (i.e., the upper 64 bits of the dividend) is zero. That is, the CPU effectively computes \code{0:rax/rcx}, and the result is exposed through the memory access at line 3.

For example, without speculation, $2^{64}/4$ and $2^{63}/2$ yield the same result. With ZDI, the former speculatively yields $0$ while the latter yields $2^{62}$. 
An attacker can observe the result from line 3, and thus learn more information about the division operands than would be possible without speculation.
Furthermore, line~3 speculatively executes an unexpected memory ready, which causes additional information leakage.

In our experiments, the predicted value was always zero, regardless of the processor state.
Moreover, even though we tested several multi-register instructions (such as \code{MUL}), we only observed value speculation in \code{DIV} and \code{IDIV}.

\vspace{0.1cm}

In the next sections, we first discuss the basic mechanism of leakage detection, and then describe the techniques used to discover speculative leaks automatically.

    \section{Background: Testing for Speculative Leaks}
\label{sec:bak}

This section provides the background on model-based testing for speculative leaks.
We start by introducing speculation contracts~\cite{Guarnieri2021}, a form of a leakage model that captures speculative microarchitectural leaks at the ISA level.
Next, we describe how contracts can be used to detect unknown speculative leaks.
We conclude by presenting an overview of Revizor~\cite{oleksenko2022revizor}, a black-box random testing tool for detecting speculative leaks, which is the basis of our implementation.

\subsection{Microarchitectural Leakage and Contracts}
\label{subsec:bak-leak}

When a CPU executes a program, its execution changes the CPU's microarchitectural state. An attacker can observe some of these changes via side channels~\cite{Tromer10,Yarom14,Osvik2006}.
We call such observable changes a \term{hardware trace}.

The process of collecting a hardware trace can be described as a function $\attack$ that takes a program $\program$, a program's input $\archstate$, and a microarchitectural context~$\microstate$, and returns a hardware trace:

\begin{equation*}
    \htrace{}=\attack{}(\program{},\archstate{},\microstate{})
\end{equation*}

A program leaks information when its hardware traces depend on the input $\archstate$, as the attacker can distinguish different inputs (potentially containing secrets) by comparing traces.

The leak could be caused not just by the normal program execution but also by the instructions executed transiently; that is, by those instructions that the CPU executes but never retires.
We call such leaks \term{speculative leaks}.
For example, the hardware trace may expose secrets accessed after a branch misprediction, as demonstrated in Spectre~\cite{Kocher2018}.

To aid the development of defences against speculative leaks, \term{speculation contracts}~\cite{Guarnieri2021} have been proposed as an abstract ISA-level specification of the expected microarchitectural leaks (i.e., a speculative leakage model).
A contract describes the information expected to leak on a CPU while executing a victim program $\program$ with an input $\archstate$.

A contract abstracts away the microarchitectural details of the function $\attack$, and provides a high-level (i.e., ISA-level) function $\contract$ that returns a trace of expected observations (\term{contract trace}): 

\begin{equation*}
    \ctrace{}=\contract{}(\program{},\archstate{})
\end{equation*}

A speculation contract annotates ISA instructions with
\begin{inparaenum}[(a)]
    \item an \emph{observation clause} that describes the information disclosed by the instruction, and
    \item an \emph{execution clause} that describes whether (and how)  instructions trigger speculation.
\end{inparaenum}

\begin{example}
    \ctseq{} is a contract that describes the leakage expected on a CPU with cache side channels and without speculative execution.
    To this end, the observation clause of \ctseq{} exposes the addresses of all memory accesses (loads and stores) and of all control-flow operations (jumps, calls, etc).
    Its execution clause is empty for all instructions, which means that no instruction is expected to trigger speculation.
\end{example}

\begin{example}
    \ctcond{} is another contract, which describes the leakage expected on a CPU with a branch predictor.
    Its observation clause is identical to \ctseq{}; its execution clause prescribes that all conditional branches always speculatively take a wrong target.
    Thus, \ctcond{} exposes the information leaked by the transient instructions that were executed due to branch prediction.
\end{example}

\subsection{Detecting Leaks via Relational Analysis}
\label{subsec:bak-detect}

Speculation contracts have been applied to detect information leaks in CPUs~\cite{oleksenko2022revizor}.
A contract violation (i.e., an \emph{unexpected} leakage) is discovered by comparing the leakage according to contract traces (i.e., the \emph{expected} leakage) with the leakage in hardware traces (i.e., the \emph{observed} leakage on the CPU under test).

\begin{definition}[\bf Violation]\label{def:ctr-sat}
    A CPU violates~\cite{Guarnieri2021} a contract $\contract$ if there exists a program $\program$, a pair of inputs $(\archstate,\archstate')$, and a microarchitectural state $\microstate$, such that $\contract(\program,\archstate)=\contract(\program,\archstate')$, and $\attack(\program,\archstate,\microstate) \neq \attack(\program,\archstate',\microstate)$.
\end{definition}

An evidence to a contract violation is called a \term{counterexample}, and it consists of a program $\program$, a pair of inputs $\archstate,\archstate'$, and a microarchitectural state $\microstate$ that match~\defref{ctr-sat}.
A counterexample is an evidence to an unexpected leakage because the attacker can distinguish $\attack(\program,\archstate,\microstate)$ from $\attack(\program,\archstate',\microstate)$, while these executions are supposed to be indistinguishable according to the contract.

\subsection{Model-based Relational Testing with Revizor}
\label{subsec:bak-revizor}

Model-based tools~\cite{Nemati2020a,buiras2021micro,oleksenko2022revizor} apply the above approach to detect unexpected leaks in CPUs.
In this paper, we base our work on one such tool, called Revizor~\cite{oleksenko2022revizor}.

Revizor searches for contract counterexamples by generating random test cases. A \term{test case} consists of a random program $\program$ and a sequence of random inputs $[\archstate_0, \archstate_1, ...]$. For each program-input combination Revizor collects the corresponding contract and hardware traces:
\begin{itemize}
    \item Contract traces are collected by the \term{contract model}, an executable version of the contract implemented with an ISA emulator (QEMU).
          The emulator is modified to record observations according to the contract observation clause, and to implement speculation according to its execution clause.
          The model collects traces by executing the program $\program$ with each of the inputs $\archstate$ on this emulator, and then retrieving the recorded observations.
    \item Hardware traces are collected by the executor.
    It implements $\attack$ by executing the program $\program$ with each of the inputs $\archstate$ on the target CPU.
    The microarchitectural state $\microstate$ is set indirectly, as it is not directly accessible on black-box CPUs: Each program execution inherently modifies the microarchitectural state, which sets $\microstate$ for the following executions. 
    Hardware traces are collected by monitoring the microarchitectural changes caused by each execution; specifically, Revizor collects traces by monitoring L1D cache state with performance counters.
\end{itemize}
After collecting the traces, Revizor checks if any of them satisfies the definition of violation (\defref{ctr-sat}). For this, it groups the inputs that produce the same contract trace into \term{equivalence classes}, and checks if all hardware traces corresponding to inputs in the same class match. If there is at least one pair of different traces (i.e., leakage according to~\defref{ctr-sat}), these constitute a counterexample to the contract, and Revizor reports the unexpected leakage to the user.

\begin{figure}[t]
    \begin{lstlisting}[keywordstyle=\texttt]
CMP rax, 10  // compare rax with 10
JNE .END     // jump if not equal
MOV rax, [rbx]  // load from address in rbx\end{lstlisting}
    \vspace{-2pt}
    \caption{A program that produces a counterexample to \ctseq{}.}
    \label{fig:counterexample-ctseq}
\end{figure}

\begin{example}\label{ex:bak-detect}
    Consider a round of a testing campaign where a CPU with branch prediction is tested against \ctseq{}.
    The round begins by generating a random program shown in~\figref{counterexample-ctseq}, and a sequence of random inputs\footnote{In practice, an input assigns values to multiple registers and to several pages of memory. This example is simplified for clarity.}:

    $\archstate_1$\code{=\{rax=10,rbx=5 \}}, $\archstate_2$\code{=\{rax=10,rbx=20\}}

    $\archstate_3$\code{=\{rax=40,rbx=10\}},
    $\archstate_4$\code{=\{rax=20,rbx=70\}}

    \noindent
    Revizor executes this test case on the contract model for \ctseq{}, which collects the jump targets (line~2) and the load addresses (line~3), producing the following traces:

    \code{ctrace$_1$=\{load *5 \space \}},
    \code{ctrace$_2$=\{load *20  \}}

    \code{ctrace$_3$=\{jump .END\}},
    \code{ctrace$_4$=\{jump .END\}}

    \noindent
    Note that \code{ctrace$_3$} and \code{ctrace$_4$} do not include the load because it is skipped when \code{rax==10}.

    Next, Revizor executes the test case on the target CPU while monitoring cache evictions with a side-channel attack, resulting in the following traces (CL stands for cache line):

    \noindent\hspace{1pt}
    \code{htrace$_1$=\{evict CL *5 \}},
    \code{htrace$_2$=\{evict CL *20\}}

    \noindent\hspace{1pt}
    \code{htrace$_3$=\{evict CL *10\}},
    \code{htrace$_4$=\{evict CL *70\}}

    In the first two traces, the branch is not taken, and the loads from addresses \code{5} and \code{20} evict the corresponding cache lines (if the cache line size is 64 bytes, they evict from the same cache set). In addition, the first two executions train the branch predictor, so the next two executions experience mispredictions. For input $\archstate_3$, the CPU speculatively executes a load from address \code{10}, and for $\archstate_4$, a load from address \code{70}. With 64-byte cache lines, $\archstate_3$ and $\archstate_4$ evict different cache lines, which results in different hardware traces.

    Accordingly, the last two inputs form a counterexample: \code{ctrace$_3=$ctrace$_4$} and \code{htrace$_3\neq$htrace$_4$}. Revizor detects it and reports to the user.
\end{example}

\myparagraph{Handling microarchitectural states}
We now provide additional details on how Revizor handles the microarchitectural state $\microstate$ when testing a program $p$ with inputs $[\archstate_0,\archstate_1,\ldots]$.
Before executing an input, Revizor partially resets the microarchitectural state $\microstate$ by flushing caches, invalidating TLB, flushing microarchitectural buffers with \code{VERW}, and pushing memory fences into the pipeline.
The rest of $\microstate$ (e.g., the state of branch predictors and of other internal buffers) is set by the execution of the test case itself: 
Since the program is executed with each of the inputs in a sequence, without interruptions, the executions of $[\archstate_0...\archstate_{n-1}]$ indirectly set $\microstate_n$ for $\archstate_n$.
Note that this mechanism is imperfect and parts of the state may not be randomized; see~\cite{oleksenko2022revizor} for further discussion of this point.

\myparagraph{Input Effectiveness}
To form a counterexample, we need two inputs with the same contract trace but different hardware traces; we call such inputs \emph{effective}.
In other words, given a program $\program$ and inputs $I = [\archstate_0, \archstate_1, ...]$, an input $\archstate \in I$ can surface leakage only if there is another input $\archstate' \in I$ (different from~$\archstate$) such that $\contract(\program,\archstate)=\contract(\program,\archstate')$.
Otherwise, if $I$ contains no input with a contract trace identical to $\archstate$, then this inputs becomes a wasted effort as it cannot, by definition, be used for leakage detection; we call it an \emph{in}effective input.

\begin{example}\label{ex:ineffective-inputs}
    Consider the program in~\figref{counterexample-ctseq}, executed with the following inputs: $\archstate_1$\code{=\{rax=0,rbx=1\}}, $\archstate_2$\code{=\{rax=1,rbx=1\}}, $\archstate_3$\code{=\{rax=0,rbx=2\}}.
    If the target contract is \ctseq{}, the first two inputs are effective: They produce the same contract trace \code{\{load *1\}}. The input~$\archstate_3$, however, produces the trace \code{\{load *2\}}, and since it is the only input to produce this trace, $\archstate_3$ is ineffective.
\end{example}

    \section{Design and Implementation}
\label{sec:design}

This section presents an approach to increase the speed of testing for speculative leaks and its implementation for Intel and AMD CPUs.

Our key observations is that random testing tools produce very few \term{effective test cases}; that is, few test cases have potential for surfacing a speculative leak.
Yet the existing tools spend just as much time on the effective cases as on the ineffective ones.
Changing this balance can significantly increase the testing speed.

We identify the following conditions that make a test case effective:
\begin{enumerate}
    \item \textit{Misspeculation}: The test case must trigger speculation, and the speculation must be based on an incorrect prediction. This will lead to transient instruction that are issued but never retire.
    \item \textit{Trace of misspeculation}: When the misspeculation is triggered, some of the transient instructions must affect the hardware trace to make a leakage observable.
    \item \textit{Effective Inputs}: To detect leakage with relational analysis, the inputs that triggers misspeculation must be effective, as described in~\subsecref{bak-revizor}.
\end{enumerate}

Next, we illustrate conditions 1 and 2 using examples (condition 3 is illustrated above in~\Cref{ex:ineffective-inputs}).
In the following, we assume a CPU that \emph{only} speculates over conditional branches and the traces are obtained via L1D cache side-channel.

\begin{example}
    The following program is ineffective because it does not meet condition~1:
    \begin{lstlisting}[style=embedded]
CMP rax, 10
MOV rax, rbx\end{lstlisting}
    The program has no instructions that would trigger speculation,  thus it cannot produce a speculative leak.
\end{example}

\begin{example}
    The following program is ineffective because it does not meet condition~2:
    \begin{lstlisting}[style=embedded]
CMP rax, 10
JNE .END
ADD rax, rbx
\end{lstlisting}
    Even if the branch (line~2) is mispredicted, the speculation will not affect the hardware trace, because the program has no memory accesses.
\end{example}

\begin{example}\label{ex:observation-filter}
    The following program is ineffective because it does not meet condition~2:
    \begin{lstlisting}[style=embedded]
CMP rax, 10
JNE .l1
MOV rax, [rbx]
.l1: MOV rax, [rbx]\end{lstlisting}
    Even though it has a conditional branch and a memory access after it, the program also has a non-speculative load (line~4), which uses the same address as the speculative one (line~3). Accordingly, whenever the program evicts a cache line speculatively, the eviction is hidden by the non-speculative access to the same address.
\end{example}

\subsection{High-level Algorithm}
\label{subsec:des-algo}

With the above conditions, we optimize the testing process to focus primarily on the effective test cases.
We identify that the most time-consuming stage of relational testing is the collection of contract traces, because it involves execution of test cases on an ISA-level emulator. Thus, our goal is to prune the test cases before reaching this stage.

\begin{figure}[]
    \centering
    \includegraphics[width=.68\linewidth]{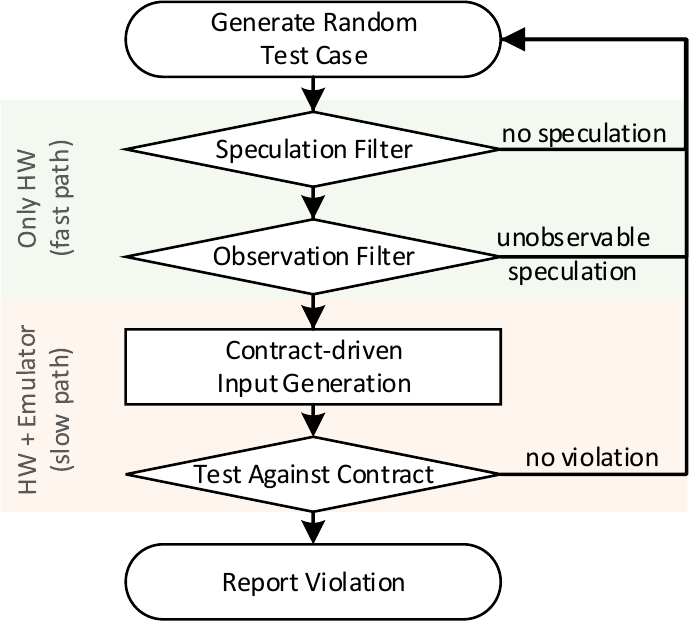}
    \caption{Main testing algorithm.}
    \label{fig:algo}
\end{figure}

We use the algorithm in~\figref{algo}.
A testing round begins by generating a random test case.
First, the \emph{speculation filter} (\subsecref{des-spec}) checks condition 1 by monitoring performance counters.
Next, the \emph{observation filter} (\subsecref{des-observ}) checks condition 2 by performing a serialization experiment.
If the test case fails any of these filters, it is discarded.
Otherwise, the test case is passed to the \emph{contract-driven input generator} (\subsecref{des-boost}), which ensures condition 3 by creating additional inputs such that every input gets at least one other input in its equivalence class.
Only then, when all three conditions are met, the test case is passed to the time-consuming relational analysis to check for contract violations.

Notably, the speculation and observation filters check the conditions conservatively, and they err on the side of permitting the test cases that produce uncertain results~(\subsecref{des-issues}).

We next describe the ideas behind each of the techniques, as well as technical details of their implementation. Our implementation is based on Revizor (described in~\subsecref{bak-revizor}).

\subsection{Speculation Filter}
\label{subsec:des-spec}

The task of the speculation filter is to find the test cases that trigger misspeculation.
Given a test case (i.e., a program $\program$ and multiple inputs $[\archstate_1,\archstate_2, \ldots]$), the filter executes the program on the target CPU with each input and monitors speculation-related performance counters in the process.
If at least one of the executions produces misspeculation (reflected in a change to the selected performance counters), then the test case is passed down to the next stages;
otherwise, it is discarded.

\myparagraph{Implementation}
On Intel CPUs, the natural candidates for detecting misspeculations are the counters called \code{MACHINE\_CLEARS.COUNT} (the number of issued machine clears) and \code{BR\_MISP\_RETIRED.ALL\_BRANCHES} (the number of branch mispredictions).
As noted by Ragab et al.~\cite{ragab2021rage}, these counters are sufficient to detect all types of speculation.

In our experiments, however, we observed that \code{MACHINE\_CLEARS.COUNT} occasionally misses a machine clear and thus, it cannot be used as a reliable signal.
Instead, our implementation relies on \code{INT\_MISC.RECOVERY\_CYCLES}, which counts recovery cycles from both machine clears and branch mispredictions, and which in practice provides more reliable results.
To double check its readings, we use another pair of counters---\code{UOPS\_ISSUED.ANY} and \code{UOPS\_RETIRED.SLOTS}---whose difference gives the number of transient (i.e., never retired) micro-operations.
Concretely, the speculation filter detects a misspeculation if either the number of recovery cycles is non-zero (i.e., \code{INT\_MISC.RECOVERY\_CYCLES} is incremented during the execution), or the number of transient micro-operations is non-zero (\code{UOPS\_ISSUED.ANY-UOPS\_RETIRED.SLOTS~>~0}).

Similarly, on AMD CPUs, we used the counters \code{PMCx0C1} (retired micro-operations) and \code{PMCx0AB} (micro-operations dispatched from the decoder).

\subsection{Observation Filter}
\label{subsec:des-observ}

\newcommand{\serialized}[1]{{#1}_\mathit{ser}}

The task of the observation filter is to find the test cases that produce observable speculation traces.
Given a program~$\program$, the observation filter first creates a serialized version of the program  $\serialized{\program}$ by injecting a serialization fence---\code{lfence}---after every instruction. Then, the filter executes both $\program$ and $\serialized{\program}$ over all inputs $[\archstate_1, \archstate_2, \ldots]$ in the test case, and collects the hardware traces. Since \code{lfence}s stop speculative execution,  differences in the hardware traces of $\program$ and $\serialized{\program}$ over an input $\archstate$ are the result of transiently executed instructions.
The filter admits a test case if $\program$ and $\serialized{\program}$ produce different traces over at least one of the inputs;
otherwise, the test case is discarded.

\myparagraph{Implementation}
Our implementation relies on Revizor's mechanism for collecting hardware traces (see~\subsecref{bak-revizor}).
This mechanism relies on a side-channel attack to observe changes in the L1D cache state.
Accordingly, the observation filter looks for differences in the L1D cache state between $\program$ and $\serialized{\program}$.
These differences are caused by transient memory accesses during $\program$'s execution that evict a cache line that is \emph{not evicted} by the (non-speculative) memory accesses in $\serialized{\program}$.\looseness=-1

\subsection{Contract-driven Input Generator}
\label{subsec:des-boost}

\newcommand{\deps}[2]{\mathit{Dep}_{#2}(#1)}
\newcommand{\instr}{\texttt{instr}}
\newcommand{\depMap}{\mathit{DMap}}
\newcommand{\readFrom}[1]{\mathit{read}(#1)}
\newcommand{\writeTo}[1]{\mathit{write}(#1)}

The task of the contract-driven input generator (CIG) is to ensure that all inputs are effective.
Random generation is rarely successful in this task, as the likelihood of two random inputs producing the same contract trace can be extremely low (e.g., in the experiment~\subsecref{boost-eval}, only 6\% of random inputs were effective).
CIG solves this problem by creating additional inputs based on the feedback from the contract model.

Given a program $\program$ and an input $\archstate$, CIG generates additional (and different) inputs $[\archstate_1, \archstate_2 \ldots]$ such that $\contract(\program,\archstate) = \contract(\program,\archstate_j)$:
\begin{enumerate}
    \item CIG tracks contract-level dependencies to compute the set $\deps{\program,\archstate}{\contract}$, which contains all dependencies of the contract trace $\contract(\program,\archstate)$.
          Concretely,  $\deps{\program,\archstate}{\contract}$ contains all parts of $\archstate$ (registers and memory locations) that may affect the trace.
    \item CIG generates new inputs $[\archstate_1, \archstate_2 \ldots]$ by mutating all the values in $\archstate$ that do not belong to $\deps{\program,\archstate}{\contract}$.
          Since parts of the input not in $\deps{\program,\archstate}{\contract}$ do not affect the contract trace, all generated inputs are contract-equivalent.
\end{enumerate}

\myparagraph{Correctness of effective input generation}
Contract-level dependency tracking guarantees that $\deps{\program,\archstate}{\contract}$ contains \emph{all} dependencies that may influence the contract trace.
As a result, for all inputs $\archstate,\archstate'$, if $\archstate$ and $\archstate'$ agree on the values of all dependencies in $\deps{\program,\archstate}{\contract}$, then $\contract(\program,\archstate) = \contract(\program,\archstate')$.
Contract-driven input generation leverages this guarantee to produce contract-equivalent inputs by mutating the parts of $\archstate$ that are not in $\deps{\program,\archstate}{\contract}$.

\myparagraph{Implementation}
We instrumented the Revizor's contract model to track dependencies between memory locations and registers at each step of program execution.
The instrumentation keeps track of a dependency map $\depMap$ that assigns to each location (a register, a memory address, or a flag) its set of dependencies, i.e., the set of initial values that might have contributed to the location's current value.
The map also tracks dependencies for the program counter \texttt{pc}.
The map is updated throughout the execution as follows:
\begin{itemize}
    \item The initial dependency map $\depMap_0$ is such that $\depMap_0(l) = \{l\}$ for each location $l$, where a location can be a register, a memory address, or a flag.

    \item For each instruction $\instr$, we compute the read set $\readFrom{\instr}$ (containing all locations read by $\instr$) and the write set $\writeTo{\instr}$ (containing all locations that $\instr$ writes to).
          Note that for control-flow instructions, the program counter \code{pc} is a part of the write set $\writeTo{\instr}$.

    \item Whenever we execute an instruction $\instr$, we update the dependency map $\depMap$ to $\depMap'$  by tracking the new dependencies from $\instr$'s read set and from the program counter to $\instr$'s write set.
          That is, for each $l \in \writeTo{\instr}$, the new dependency set $\depMap'(l)$ is $\depMap(\texttt{pc}) \cup \bigcup_{l \in \readFrom{\instr}} \depMap(l)$.
    \item Whenever the contract model explores a speculative path, we create a copy of the dependency map which is used (and updated) throughout the speculative transaction (similarly to how the contract model tracks the speculative program state).
          When the speculative transaction terminates, the speculative dependency map is discarded and the old dependency map is restored.
\end{itemize}

We compute the set $\deps{\program,\archstate}{\contract}$ of dependencies associated with the trace $\contract(\program,\archstate)$ as follows.
Whenever the contract produces an observation $o$ and the current dependency map is $\depMap$, we compute the locations $L$ that influence $o$'s values and we update $\deps{\program,\archstate}{\contract}$ by adding the program counter's dependencies $\depMap(\texttt{pc})$ and the dependencies of $L$'s locations $\bigcup_{l \in L} \depMap(l)$.
For instance, if the observation exposes the accessed memory address, then the set $L$ contains the operands determining the address.

\begin{example}\label{ex:dt-depmap}
    Consider the following program:
    \begin{lstlisting}[style=embedded]
CMP rax, 10
JNE .l1
MOV rax, [rbx]
.l1: MOV rbx, [rax]\end{lstlisting}
    It is executed with the input $\archstate$\code{=\{rax=20,rbx=5\}}, and the target contract is \ctseq{}.
    During dependency tracking, the dependency map is updated as follows (here, $\depMap_j$ denotes the map after executing line $j$):

    Initially, the map $\depMap_0$ is such that each location $l$ is mapped to $\depMap_0 = \{l\}$: e.g.,

    \centerline{$\depMap_0(\code{rax})=\{\code{rax}\}$}

    \noindent
    At line 1: The comparison sets the dependencies of all flags to those of \code{rax} and of \code{pc}, since all instruction read \code{pc} by default.
    For example, the resulting dependencies of \code{ZF} are:

    \centerline{$\depMap_1(\code{ZF})=\{\code{pc,rax}\}$}

    \noindent
    At line 2: Jump propagates the dependencies of \code{ZF} to \code{pc}:

    \centerline{$\depMap_2(\code{pc})=\{\code{pc,rax}\}$.}

    \noindent
    (Line 3 is not executed for this input.)

    \noindent
    At line 4: The dependency set $\deps{\program,\archstate}{\text{\ctseq}}$ is updated, because loads are exposed in \ctseq{}. Specifically, the load address depends on \code{rax} and on \code{pc} (by default), therefore:

    \centerline{$\deps{\program,\archstate_1}{\text{\ctseq}}=$}
    \centerline{$=\depMap_2(\code{pc}) \cup \depMap_2(\code{rax}) = \{\code{pc,rax}\}$}

    After line 4, the execution finishes, and the resulting dependency set is passed down to the input generator.

    The generator takes the input $\archstate_1$, and randomly mutates those parts of it that are not included into $\deps{\program,\archstate}{\text{\ctseq}}$. In this (simplified) example, the input consists of only two registers \code{rax} and \code{rbx}, and only \code{rbx} is not included in $\deps{\program,\archstate}{\text{\ctseq}}$. Accordingly, the generator mutates \code{rbx} and leaves \code{rax} unchanged. The resulting input $\archstate'$ is:

    \centerline{$\archstate'$\code{=\{rax=10,rbx=70\}}}

    The input pair $i,i'$ is effective because they produce the same contract trace. Should speculation occur during the measurement on the target CPU, the (speculative) load at line~3 will access different addresses for inputs $i$ and $i'$, thereby surfacing a speculative leak. 
\end{example}

\subsection{Practical Issues}
\label{subsec:des-issues}

The filtering techniques described in~\subsecref{des-spec} and~\subsecref{des-observ} rely on empirical measurements, which are inherently imprecise on a black-box CPU.
Such imprecision occasionally leads to false negatives and false positives.

\begin{figure*}[t]
    \centering
    \includegraphics[width=.83\textwidth]{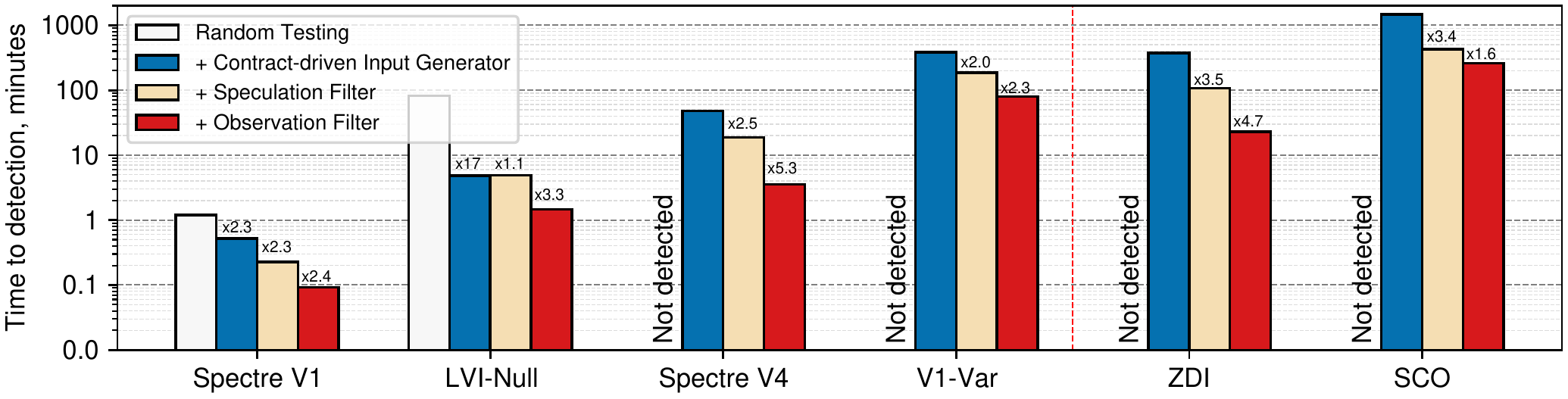}
    \caption{Detection Time: The average amount of time required to detect a leak. The numbers above the bars are the ratio between the given bar and the one left from it, which corresponds to the relative speedup between the measurements.
    }
    \label{fig:detection}
\end{figure*}

\myparagraph{False negatives}
Both speculation and observation filters implicitly assume that speculative execution is deterministic. This assumption is not always true on a black-box CPU, where we cannot directly control the microarchitectural state. Even if the test case contains the necessary instructions and data to trigger speculation, the microarchitectural state may prevent a misspeculation (e.g., if a branch predictor is not trained). Accordingly, the speculation filter may discard an effective test case because a predictor did not produce a misprediction, or the observation filter may discard the test case because the speculation was not long enough to leave a trace.\looseness=-1

Fortunately, such cases are rare. We tested both filters on several instances of known leaks, and they produced correct results. We further evaluate false negatives in detail in~\subsecref{eval-gray}.

\myparagraph{False positives}
The observation filter occasionally produces false positives that are caused by external noise. When the filter executes $\program$ and $\serialized{\program}$ with the same input and collects the corresponding hardware traces, one of the traces could get corrupted by the measurement noise. This will lead to an apparent mismatch between the traces, even though it is not caused by transient instructions.
These false positives result in some ineffective test cases not being discarded early on.
The impact is minor, however, since we observed a corrupted measurement on average only once in a million executions.

    \section{Evaluation}
\label{sec:eval}

This section evaluates the impact of the speculation and the observation filters, and of the contract-driven input generator.

We evaluate the impact across three metrics:
\emph{Detection time} is the time required to detect a leak; we consider it the key metric for our leakage-detection tool.
\emph{Testing speed} is the amount of time required to test $N$ test cases.
\emph{Detection rate} is the number of contract violations detected within $N$ test cases.
Detection time is a compound metric, as it depends on both the testing speed and the detection rate.

In~\subsecref{eval-speed}, we measure the impact of our techniques on the detection time. 
Then, we break it down into its components. In~\subsecref{eval-gray} we measure the changes in the testing speed and the detection rate caused by speculation and observation filters; and in~\subsecref{boost-eval}, we evaluate the changes caused by CIG.

\subsection{Experimental Setup}

\begin{figure*}[]
    \centering
    \includegraphics[width=.85\textwidth]{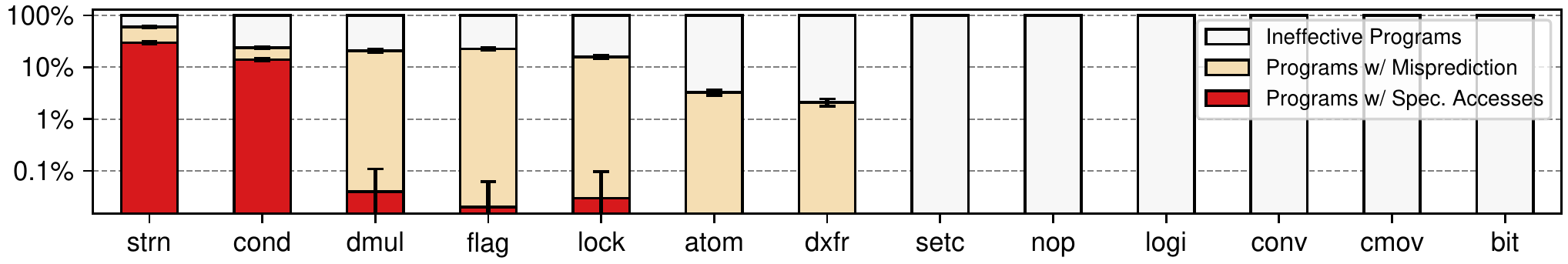}
    \vspace{-6pt}
    \caption{Share of random test cases with misspeculation and speculative memory accesses in different x86 subsets. The subsets are described in~\appref{subsets}.}
    \label{fig:gains-illustration}
\end{figure*}

We perform the experiments on an Intel Xeon E-2288G CPU\footnote{We also performed the measurements on the other machines from~\tabref{results1}, but the results were very close to Intel Xeon, hence we do not present them. The only notable difference was in~\figref{gains-illustration} where the AMD CPU did not show any signs of speculation in \code{dmul} and \code{flag}.} with all microcode patches against speculative leaks enabled; the only exception was the V4 target in~\subsecref{eval-speed}, where the corresponding patch was disabled.

All the experiments used the same configuration of the test case generator (unless mentioned otherwise): program size---32 instructions; number of memory accesses per program---8; input generation entropy---16 bits.

\subsection{Detection Time}
\label{subsec:eval-speed}

\begin{figure*}[]
    \centering
    \includegraphics[width=.85\textwidth]{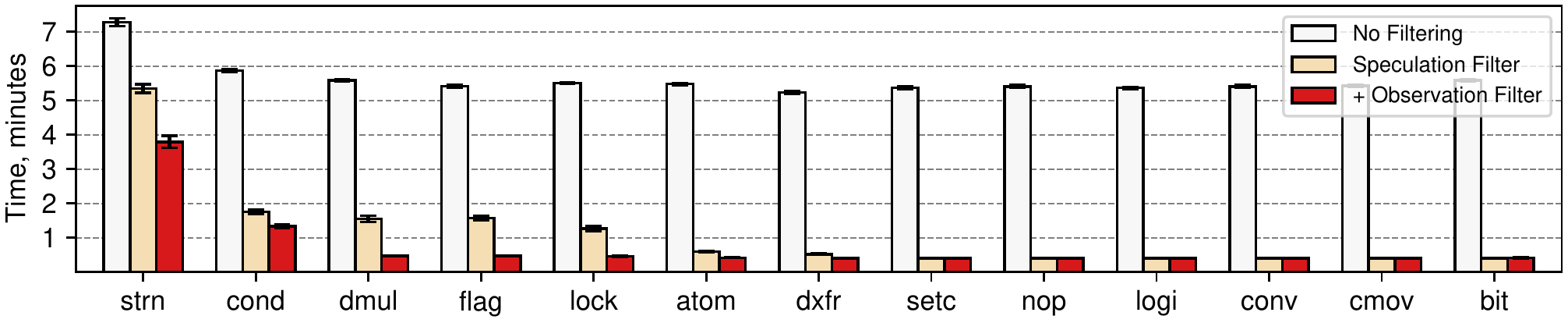}
    \vspace{-6pt}
    \caption{Change of the testing speed caused by speculation and observation filtering, for subsets of x86 ISA. The subsets are described in~\appref{subsets}.}
    \label{fig:fractions-perf}
\end{figure*}

Our first research question is: \emph{Do our techniques reduce the leak detection time?}

The answer is affirmative, yet the extent of the reduction depends on the leak. To evaluate it, we consider Spectre V1, V4, and LVI, a variant of V1 (called V1-Var), as well as the new speculative leaks---ZDI and SCO---described in~\secref{findings}.
We perform the evaluation on several testing configurations (described in detail in~\appref{configs}), which we specifically pick such that our tool detects only one type of leaks in each configuration, and any other leaks are unlikely to surface.

For example, in the configuration ``Spectre V1'', we test a combination of subsets \code{nop}, \code{bit}, \code{cond}, \code{cmov}, \code{conv}, \code{dxfr}, \code{flag}, \code{setc}, \code{logi} (see~\appref{subsets}) against the \ctseq{} contract. This configuration can surface V1 because the instruction set contains conditional branches, but it cannot surface the other leaks because the firmware patch is enabled (prevents V4), microcode assists are not permitted (prevents LVI), the subsets do not include divisions (prevents ZDI) and string operations (prevents SCO).

For each of these configurations, we perform an experiment where we execute 100'000 test cases with 100 inputs each; for ZDI and SCO, we execute 500'000 test cases (these leaks are harder to detect, so we need more test cases).
We measure the overall execution time, count the number of detected contract violations, and hence calculate the average time to violation.

Each experiment is performed with each of our techniques:
\begin{inparaenum}[(i)]
    \item fully random testing (baseline),
    \item with contract-driven input generation,
    \item with speculation filtering, and
    \item with observation filtering.
\end{inparaenum}
We fix the seed for random generation to ensure that the different techniques are evaluated on the exact same sequence of test cases.

\myparagraph{Results}
\figref{detection} shows the results. Note that the vertical axis is in log-scale.
We highlight the following findings:
\begin{asparaitem}
    \item In {\em isolation}, speculation filtering, observation filtering, and contract-driven generation already lead to significantly faster detection of almost all leaks. In {\em combination}, they can reduce the detection time by a factor of 10--50.
    \item An exception is the configuration LVI-Null, where the speculation filter had almost no impact on the detection time. This is because this speculation type is easy to trigger: any access to a page with an unset Dirty bit suffices. Thus, most of the test cases experienced a machine clear, and the speculation filter was not useful.
    \item Without the techniques presented in this paper (i.e. in the baseline experiment), only V1 and LVI-Null are detected. V4 and ZDI are not detected because the effectiveness of random inputs is particularly low and speculation is hard to trigger, which is why contract-driven generation is instrumental for detection. Detection of SCO requires a pair of inputs such that in one of the inputs the strings match and in the other do not; finding a randomly-generated pair that both matches this description \emph{and} is effective, is very unlikely. V1-Var requires a microarchitectural race condition, which is hard to trigger with random programs.
\end{asparaitem}

\myparagraph{Sources of lower detection time}
To investigate the source of such a drastic improvement in the detection time, we   break it down into its components.
Detection time is effectively a ratio between testing speed and detection rate.
Our techniques impact these parameters differently:
Speculation and observation filters improve the testing speed (because they discard ineffective test cases), but they also decrease the detection rate (because of false negatives, see~\subsecref{des-issues}).
On the other hand, contract-driven generation increases the detection rate  (because all inputs can be used for leakage detection), but decreases the testing speed (because dependency tracking takes additional time).
We evaluate these effects in~\subsecref{eval-gray} and~\subsecref{boost-eval}.

\subsection{Impact of Speculation and Observation Filters}
\label{subsec:eval-gray}

\myparagraphnodot{Testing speed:}
\emph{How much does speculation and observation filtering change the testing speed?}

The answer to this question depends on the instructions from which test cases are randomly generated: they govern the likelihood of triggering speculation and hence the effect of filtering out test cases that neither speculate nor perform speculative memory accesses.

We perform experiments on the subsets of the x86 ISA from~\Cref{tab:results1}.
For each subset, we generate 1000 test cases with 100 inputs each, and we measure (i) the fraction of test cases that passes each filter, and (ii) the time required for processing all test cases with none, one, or both filters enabled. The measurement is repeated 10 times with different seed values of the test case generator. We report the mean values and the standard deviations.

\myparagraph{Results} The results for (i) are given in~\figref{gains-illustration}, for (ii) in~\figref{fractions-perf}. We highlight the following findings:
\begin{asparaitem}
    \item Speculation and observation filtering consistently improve the testing speed for all subsets of the x86 ISA we considered (\figref{fractions-perf}). The reason for this improvement is that most of the test cases experience no speculation, and neither do they trigger speculative memory accesses (\figref{gains-illustration}), which is why pruning them early is beneficial.
    \item The largest speedups of an order of magnitude are achieved in the ISA subsets that {\em never} experience speculation (e.g., \code{cmov}, \code{nop}) because virtually all test cases are filtered out. The smallest speed-ups are found in the instruction sets that have a true violation (e.g., \code{cond} could trigger Spectre V1). There speculation is ubiquitous, and a large portion of the test cases passes the filter. (A similar behavior is observed if the contract permits speculation, see~\subsecref{dis-limit}.)\looseness=-1
    \item The impact of false positives on speculation filtering (\subsecref{des-spec}) is noticeable in \code{flag} and \code{lock}, see \figref{gains-illustration}: Many ineffective test cases pass the first filter, which is why the speed-up from speculation filtering is less significant. However, these cases do not pass through the observation filter, so they have almost no effect on the testing speed when both filters are enabled.
    \item Notably, the baseline testing speed was somewhat different between the instruction sets, because certain instructions take longer to execute on the contract model. For example, string operations (\code{strn}) produce many memory accesses, which slows down contract tracing.
\end{asparaitem}

\myparagraphnodot{Detection rate:}
\emph{How many contract violations are missed due to false negatives of filtering?}

\begin{figure}[tb]
    \center
    \footnotesize
    \begin{tabular}{l|c|c|c}
                   & No        & Speculation & + Observation \\
                   & Filtering & Filter      & Filter        \\
        \hline
        Spectre V1 & 1577      & 1585        & 1537          \\
        Spectre V4 & 15        & 15          & 14            \\
        LVI-Null   & 167       & 168         & 159           \\
        V1-Var     & 3         & 3           & 3             \\
        \hline
        ZDI        & 11        & 11          & 11            \\
        SCO        & 3         & 4           & 4             \\
        \hline
    \end{tabular}
    \caption{Number of detected violations within a batch of test cases, tested with different filtering levels. In the first 4 rows, the batch size is 100k cases; in the last 2, the size is 500k.}
    \label{tab:filter-fp}
\end{figure}

The inherent limitation of speculation and observation filtering  is that some of the effective test case could be mistakenly discarded (see~\subsecref{des-issues}). We seek to estimate the amount of such wrongly discarded cases by  comparing the numbers of violations detected in the first experiment~\subsecref{eval-speed}.

\tabref{filter-fp} presented the number of violations detected within the same sequence of test cases, but with different filtering levels.
As we see, the number of violations changed very little when we applied the filtering techniques. It implies that the rate of false negatives was low and the techniques had little impact on the detection rate. Curiously, in some of the measurements, more violations were detected with filtering rather than without. We attribute it to the instability of speculation in some of the test cases, which introduces a measure of randomness into this experiment.

\subsection{Impact of Contract-driven Input Generation}
\label{subsec:boost-eval}

\myparagraphnodot{Detection rate:}
\emph{How many more contract violations are found due to contract-driven generation?}

While random input generation may result in ineffective inputs, contract-driven input generation (CIG) produces effective inputs by construction (\subsecref{des-boost}).
Thus, each test case is tested more thoroughly with CIG-produced inputs, and we expect to detect more contract violations.

We check if this expectation by looking at the total number of violations detected in~\subsecref{eval-speed}, shown in~\tabref{contract-detect}.
The columns \emph{Random Testing} and \emph{CIG} are the number of violations detected with the corresponding input generation method, given the same sequence of programs.
With random generation, there are only a few violations, and some leaks are not detected at all.
Meanwhile, with CIG, the number of violations increases drastically, which indicates that CIG helps in finding leaks.

\begin{figure}[tb]
    \center
    \footnotesize
    \begin{tabular}{l|c|c}
                   & Random Input & Contract-driven Input \\
                   & Generation   & Generation            \\
        \hline
        Spectre V1 & 272          & 1577                  \\
        Spectre V4 & 0            & 15                    \\
        LVI-Null   & 4            & 167                   \\
        V1-Var     & 0            & 3                     \\
        \hline
        ZDI        & 0            & 11                    \\
        SCO        & 0            & 3                     \\
        \hline
    \end{tabular}
    \caption{Number of detected violations within a batch of programs, where their inputs are generated either randomly or with CIG. In the first 4 rows, the batch size is 100k programs; in the last 2, the size is 500k.}
    \label{tab:contract-detect}
\end{figure}

To illustrate the source of such a significant improvement, we perform an additional experiment that shows just how ineffective random generation is.
For this, we randomly generate 1000 programs and for each of them we generate 100 random inputs.
For each program, we execute the inputs on the \ctseq{} contract model, and we count the number of effective inputs (i.e., those for which there is at least another input producing the same contract trace).
For \ctseq{}, the likelihood of randomly generating an effective input depends on  the number of memory accesses and of control-flow paths.
Thus, we repeat the experiment in four configurations: with 1 and 4 conditional branches, and with 4 and 16 memory accesses.\looseness=-1

\begin{figure}[]
    \center    \footnotesize
    \begin{tabular}{l|c|c|c|c}
        Configuration    & BB=1  & BB=1   & BB=4  & BB=4   \\
                         & Mem=4 & Mem=16 & Mem=4 & Mem=16 \\
        \hline
        Effective inputs & 4.5\% & 0\%    & 5.7\% & 0\%    \\
        \hline
        \hline
    \end{tabular}
    \vspace{3pt}
    \caption{Share of effective inputs produced by random input generation, for different program generator configurations. \emph{BB} stands for the number of basic blocks in a program; \emph{Mem} stands for the number of memory accesses.\looseness=-1}
    \label{tab:effectiveness}
\end{figure}

The results are in~\tabref{effectiveness}.
The share of effective inputs produced by the random generator is very low (less than 6\% in our results).
Moreover, the number of effective inputs decreases with the increase of memory accesses in the program (so, when the contract trace exposes more information).
These results indicate the key limitation of random generation for relational testing, which CIG overcomes.

\myparagraphnodot{Testing speed:}
\emph{How much does CIG cost in terms of the testing speed?}

\begin{figure}[]
    \centering
    \includegraphics[width=.85\columnwidth]{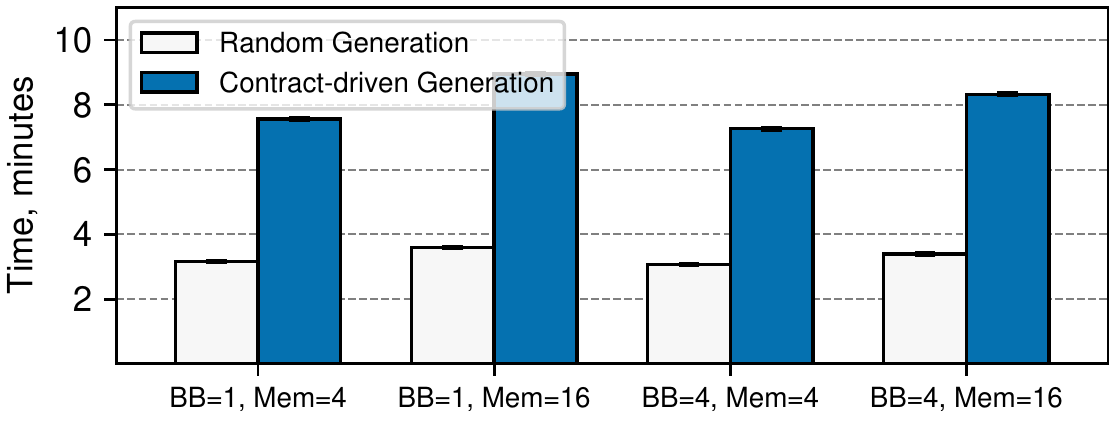}
    \vspace{-5pt}
    \caption{Performance impact of contract-driven input generation, in different program generator configurations. \emph{BB} stands for the number of basic blocks in a program; \emph{Mem} stands for the number of memory accesses.}
    \label{fig:effectiveness}
\end{figure}

Contract-driven generation, with its contract-level dependency tracking, is considerably more time-consuming than random generation, as the following experiment shows.

For each of the configurations from the previous experiment (\Cref{tab:effectiveness}), we execute 1000 test cases, each with 100 inputs, and we measure the execution time. We repeat the measurement with random input generation and with CIG.

The results are in~\figref{effectiveness}.
As we can see, CIG has a noticeable cost:
It increases the execution time (i.e., reduces the testing speed) by over 2x.
Nevertheless, this cost is acceptable, because CIG allows us to detect many more contract violations. As we saw in~\figref{detection}, even with this high cost, CIG produces a net benefit w.r.t. the detection time.\looseness=-1

\subsection{Dependency Tracking vs Symbolic Execution}\label{subsec:eval-symex}

Effective inputs can be generated using different techniques.
For instance, our contract-driven input generator (\subsecref{des-boost}) relies on dependency tracking whereas Scam-V~\cite{Nemati2020a} (another model-based testing tool) uses symbolic execution.
Here, we compare these two techniques.

\begin{figure}[]
    \centering
    \includegraphics[width=.85\columnwidth]{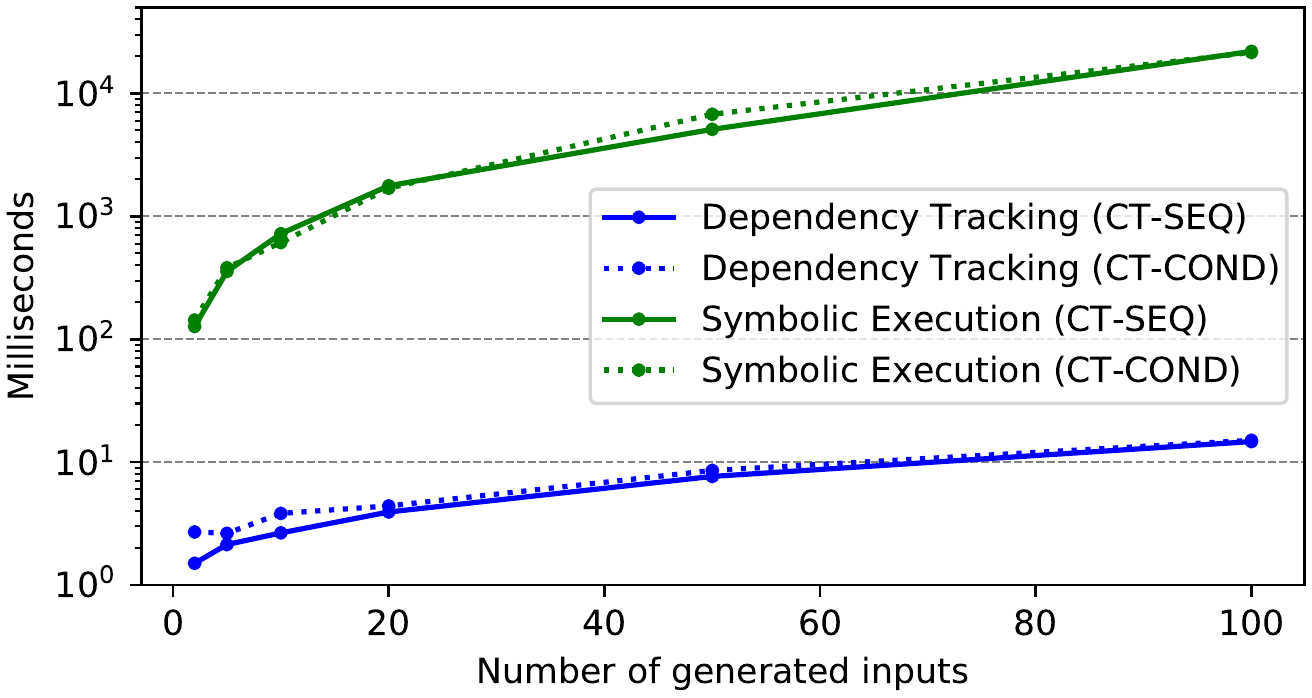}
    \caption{
        Execution time of different input-generation techniques---contract-driven input generation (\subsecref{des-boost}, in blue) and symbolic execution (in green)---for the \ctseq{} and \ctcond{} contracts.
    }
    \label{fig:symex-vs-dt}
\end{figure}

\myparagraph{Generating contract-equivalent inputs with symbolic execution}
Directly comparing the input generator from \subsecref{des-boost} with Scam-V symbolic execution approach is impossible because Revizor targets x86 whereas Scam-V targets ARM.
Instead, we implemented a contract-driven input generator based on symbolic execution by extending \textsc{Spectector}~\cite{Guarnieri2018, Fabian2022}.
This is a state-of-the-art symbolic analysis tool for verifying the absence of leaks in x86 programs with respect to the \ctseq{} and \ctcond{} contracts.
Given a program $\program$  and concrete input $\archstate$, we generate the fresh $j$-th contract-equivalent input by keeping track of the inputs $[\archstate_1, \ldots, \archstate_{j-1}]$ already generated and executing a symbolic query asking a new input $\archstate'$ such that (a) $\archstate$ and $\archstate'$ produce the same contract traces (i.e., $\contract(\program,\archstate) = \contract(\program,\archstate')$) and (b) $\archstate'$ is different from $\archstate$ and from all previously generated inputs  $[\archstate_1, \ldots, \archstate_{j-1}]$.
If the symbolic query is satisfiable, \textsc{Spectector} generates a fresh contract-equivalent input.
We use \textsc{Spectector}'s symbolic relational reasoning capabilities, which support expressing symbolic constraints over pairs of executions, to encode the aforementioned symbolic query.

\myparagraph{Experiment}
To compare the cost of CIG and symbolic execution, we measure the time needed to compute $n$ contract-equivalent inputs $[\archstate_1, \ldots, \archstate_n]$, with $n$ taking values in $\{2, 5, 10, 20, 50, 100\}$, given a program $p$ and a concrete initial input $\archstate_0$.
For each value of $n$, we generate 100 random programs with 10 concrete initial inputs each and we measure the average input generation time for both approaches.
We repeat these experiments for two different contracts: CT-SEQ and CT-COND.
The generated programs consist of  24 instructions with 1 basic block and 12 memory accesses. 

In all experiments, the symbolic memory initially allocated for \textsc{Spectector}'s inputs is limited to 20 bytes (rather than the full memory pages used by CIG).
This was necessary to limit the symbolic analysis' execution time. 

\myparagraph{Results}
\figref{symex-vs-dt} depicts the average input generation time for CIG (with dependency tracking) and symbolic execution.
We highlight the following findings:\looseness=-1
\begin{asparaitem}
    \item The input generation time increases (for both techniques) with the number of inputs to generate.

    \item Generating contract-equivalent inputs using dependency tracking is significantly faster than with symbolic execution.
    For \ctseq{}, for instance, CIG takes between 1.5 ms ($n=2$) and 14.6 ms ($n=100$) on average, whereas symbolic execution takes between about 142.7 ms ($n=2$) and 22.15 s ($n=100$).

    \item Generating inputs for the \ctcond{} contract is slightly more expensive than for the \ctseq{} contract, for both techniques.
    This is due to the COND execution clause resulting in the execution of more instructions than SEQ.

    \item Increasing the size of the generated programs by duplicating the number of instructions and memory accesses (not shown in \figref{symex-vs-dt}), results in an increase in the input generation time for both techniques.
    For \ctseq{}, CIG takes between 2.3 ms ($n=2$) and 15.6 ms ($n=100$) on average, whereas symbolic execution takes between about 183.4 ms ($n=2$) and 42.2 s ($n=100$).

\end{asparaitem}

    \section{Discussion}
\label{sec:dis}

\subsection{Impact of New Speculative Leaks}

To understand the potential security impact of SCO and ZDI, we counted the corresponding instructions in three common applications:
\begin{inparaenum}[(i)]
    \item In Linux kernel v5.14 we found 320 \code{CMPS}, 276 \code{SCAS}, and 28 64-bit divisions.
    \item In GLibC v2.27 we found 131 \code{CMPS}, 17 \code{SCAS}, and 4 64-bit divisions.
    \item In OpenSSL v1.1.1 we found 18 \code{CMPS}, no \code{SCAS}, and 2 64-bit divisions.
\end{inparaenum}
Even though we do not know how many of these instances are exploitable, the numbers show us that both SCO and ZDI could potentially affect critical applications and libraries.
We leave the investigation of software vulnerabilities caused by SCO and ZDI to future work, since the goal of this paper is to present the techniques for finding speculative leaks rather than ways to exploit them.

\myparagraph{Mitigation}
SCO can be (partially) patched at the software level by adding an \code{lfence} before the string comparison, which shortens the speculation window, although it does not prevent the speculation entirely.
ZDI can be patched by inserting an \code{lfence} after 64-bit division operations, thereby stopping the speculative execution.
At the time of writing, Intel and AMD have not announced any hardware/firmware mitigations for SCO or ZDI.

\subsection{Filtering vs Targeted Generation}

Our strategy for improving the effectiveness of testing relies on generating test cases randomly and then filtering out the ineffective ones.
Alternatively, one could target program generation towards test cases that are effective by design.
For example, adding a load after every instruction in a test case would trivially address Condition 2 in~\secref{design}.

We refrained from this approach because targeted generation can have unintended consequences:
Every speculative leak has its own unique set of requirements, and unknown leaks may have unknown requirements (e.g., ZDI requires a division with a dividend larger than $2^{64}$).
Targeted program generation that satisfy Conditions 1 and 2 might violate these unknown requirements, thereby making it less likely (or even impossible) to detect unknown leaks.
For example, by adding a load after every instruction, we increase the probability that a speculative value produced by LVI or by V4 will be overwritten, and it can reduce the chances of detecting these leaks.

\subsection{Limitations}
\label{subsec:dis-limit}

\myparagraph{Not detected leaks}
There are known leaks that have not been detected in our experiments (e.g., Meltdown~\cite{Lipp2018}, FPVI~\cite{ragab2021rage}).
The main reason is coverage: We tested a small subset of the x86-64 ISA, and most of the non-detected leaks require instructions/events outside of this subset.
For example, FPVI requires floating-point or SIMD instructions, and Meltdown requires page faults that were intentionally avoided during our experiments.
To increase the coverage and to detect such leaks, we would have to enhance the test case generator to produce more complex instructions and to handle faults.

The only exceptions (that we known of) are Straight Line Speculation (SLS)~\cite{GRSec22} and Phantom JMPs~\cite{Wikner2022}.
Our testing campaign covered all the instructions necessary for these leaks, yet they were not detected.
The reason behind it is imperfect randomization of the microarchitectural state:
These leaks require mistraining of the BTB by executing diverse (e.g., random) binaries.
However, Revizor executes each test case multiple times, which means that the BTB state is not randomized enough, resulting in not exploring such “mispredicting” BTB states.
We consider solving this issue as future work.

\myparagraph{Detection of non-speculative leaks}
Even though speculation and observation filters are instrumental in increasing test effectiveness for detecting speculative leaks, they reduce the chances of detecting non-speculative leaks (e.g., those described by Sanchez Vicarte et al.~\cite{Rodrigo2021}).
Indeed, a test case that detects a non-speculative leak might be discarded because it does not leave traces of misspeculation (i.e., it does not meet conditions 1 or 2 from~\cref{sec:design}).

\myparagraph{Dependency tracking}
Our dependency tracking is coarse-grained: it only tracks which locations influence the trace, but it does not track the relationships between locations and trace observations.
For instance, if an observation $o$ depends on the sum of registers \texttt{rax} and \texttt{rbx} being greater than 10, our dependency tracking determines that \texttt{rax} and \texttt{rbx} influence the trace.
Hence, CIG will not mutate the values of \texttt{rax} and \texttt{rbx}.
However, any mutation satisfying \texttt{rax+rbx>10} would not change the trace.
Finer-grain (but more expensive) dependency analyses might capture some of  these relationships and allow more input mutations.

\myparagraph{Testing against contracts allowing speculation}
The performance improvement of speculation and observation filters can be reduced by testing against contracts (like \ctcond{}) that allow selected speculative leaks.
Indeed, some of the tests that pass our filters (i.e., they leave traces of misspecultion) might simply represent leaks that are allowed at contract-level.
In our testing campaign, we avoided this issue by splitting the tested instruction set into subsets, which we then tested against the simple \ctseq{} contract (that forbids all speculative leak).

\subsection{Generality}

\myparagraph{Speculation and observation filters}
Our filtering techniques can be applied to any testing approach for detecting speculative leaks using randomly generated programs.
The benefits of filtering, however, depend on the specific testing approach:
As our results show, early filtering significantly improves the effectiveness of model-based testing for black-box CPUs by preventing ineffective test cases to be executed on the contract model.
In other settings (e.g., white-box testing of RTL processor designs), the execution of the contract model is unlikely to be the  performance bottleneck, so the impact of filtering might be reduced.

\myparagraph{Implementing filters for other architectures}
We implemented the filtering techniques for Intel and AMD processors using (a) performance counters for reliably detecting speculation (for the speculation filter), and (b) instructions that reliably stop speculation (for the observation filter).
Regarding point (a), other architectures provide counters tracking the effects of speculative execution.
For instance, ARM provides the \texttt{BR\_MIS\_PRED} counter tracking the number of retired mispredicted branch instructions.
Additionally, ARM also supports the \texttt{OP\_RETIRED} and \texttt{OP\_SPEC} counters counting respectively the number of retired and speculatively executed microoperations.
Regarding point (b), many other architectures have instructions that act as speculation barrier, e.g., \texttt{SB} for ARM.

\myparagraph{Contract-driven Input Generation}
CIG can be integrated into relational testing approaches for detecting leaks with respect to a given leakage contract.
In this context, CIG can be a lightweight, faster alternative to input generation techniques based on symbolic execution (\Cref{subsec:eval-symex}).
Note that CIG is not restricted to the detection of speculative leaks.
For instance, applying CIG to the \ctseq{} contract can generate inputs witnessing (architectural) leaks due to variable-timing instructions.

\myparagraph{Filters and hardware traces}
Revizor obtains hardware traces through cache side-channel attacks on the L1D cache, so this is what we used in our evaluation (\secref{eval}).
While our filtering techniques are largely agnostic of how hardware traces are obtained, different kinds of hardware trace may influence the outcome of the filters.
For instance, if hardware traces only expose whether a block is currently in the L1D cache (as currently implemented in Revizor), then the observation filter rejects the program in  \Cref{ex:observation-filter}.
In contrast, hardware traces exposing information about cache metadata (like the age of each block~\cite{Xiong2021,Vila2020}) may result in \Cref{ex:observation-filter} being accepted by the observation filter.

\myparagraph{Distributed testing}
Given that Revizor's testing process is random and the generated tests are independent, the performance issues of contract modeling could be addressed by distributing the workload.
Indeed, both the model and the executor could be distributed to an arbitrary number of nodes with a central node analyzing traces.
The filters proposed in this paper could be applied within such an architecture too.
We consider a distributed implementation as future work.

    \section{Related work}

For an overview and taxonomy of speculative attacks see~\cite{canella2019sokattacks} and the references provided throughout the paper. For an overview of software-based defenses, see~\cite{cauligi2021sok}. Here we focus on work related to automatic detection of speculative leaks and side-channels in hardware.

\myparagraph{Post-manufacture Detection of Speculative Leaks}
We begin with the tools targeting fabricated CPUs.

Revizor~\cite{oleksenko2022revizor} provides the baseline for the techniques presented in this paper. Revizor is based on relational testing against a leakage model, and it relies on random unguided generation. In contrast, speculation and observation filters use hardware signals to prune ineffective test cases, whereas contract-driven input generation uses the contract feedback to produce effective inputs. 

Scam-V~\cite{Nemati2020a,buiras2021micro} also relies on relational model-based testing, which is implemented by instrumenting test cases to record observations.
It targets ARM ISA, and it uses symbolic execution for generating inputs with identical leakage.
In contrast, we target x86, and we rely on a lightweight approach for input generation based on dependency tracking (see~\subsecref{eval-symex}). 

Transynther~\cite{Moghimi2020a} is a tool that can detect variants of microarchitectural data sampling (MDS) attacks. To this end, it fills CPU buffers with buffer-specific nonces, mutates and executes code-templates that are known to trigger speculation, and it tries to detect leakage of the nonces. In contrast, our approach searches for violations of an abstract leakage model, which enables the discovery of entirely new classes of speculation.

SpeechMiner~\cite{Xiao2020} is a framework to quantitatively evaluate CPUs for their vulnerability to known kinds of speculative attacks. The framework is scriptable and enables users to combine different kinds of primitives used in known attacks. In contrast, our tool generates fully random programs, which enables discovery of new leaks.

Ragab et al.~\cite{ragab2021rage} and Li et al.~\cite{Li2022} use counters of machine clears to discover new types of speculative leakage, albeit in a manual fashion. Speculation filtering is inspired by the idea of using machine clear counters, and it uses the counters to prune test cases during automatic analysis.

\myparagraph{Pre-silicon Detection of Speculative Leaks}
The methods for white-box pre-silicon detection speculative leaks rely on access to a detailed CPU design.

CheckMate~\cite{Trippel2018} is a tool that synthesizes exploits against microarchitectural leaks such as Spectre and Meltdown. It relies on a specification of the microarchitecture in terms of happens-before relations. In contrast, our approach relies only on minimal signals from the hardware and can be applied to commercial off-the-shelf CPUs.

IntroSpectre~\cite{Ghaniyoun2021} is an RTL-based framework to detect Meltdown-type leaks. It relies on inserting nonces into a security domain, executing randomly mutated leakage templates, and searching for the nonces in other security domains. In contrast, our approach uses relational analysis and can identify information flows beyond direct data transfer.

Fadiheh et al.~\cite{Fadiheh2019} propose to detect covert channels by model-checking an RTL model of a RISC-V CPU for violations of non-interference.

\myparagraph{Detection of Side-channels}
Several approaches automatically detect side-channels, without considering speculation.

Osiris~\cite{Weber2021} discovers novel non-speculative side-channels. For this, it generates random code snippets and measures if their execution time is secret-dependent. In contrast, we specifically focus on speculative leaks.

ABSynthe~\cite{Gras2020} creates leakage maps capturing interactions between different x86 instructions with SMT enabled, which it uses to synthesize attacks on a given target program.\looseness=-1

In principle, the result of these tools could be used to refine the hardware traces used by our proposed techniques.

\myparagraph{Leakage models and information flow checking}
The work presented in this paper speeds up testing CPUs against leakage models.
Specifically, we rely on a model for speculative leakage that is formalized as {\em speculation contracts}~\cite{Guarnieri2021}. There are alternative leakage models emerging, for example those based on ideas from weak memory models~\cite{leon22catsspectre,mosier22axiomatic}.

SecVerilog~\cite{secverilog} is a hardware design language with information-flow control built in. It uses static type checking to ensure that a system complies with an information-flow control policy by design. Connecting these policies to ISA-level leakage model~\cite{Guarnieri2021,leon22catsspectre,mosier22axiomatic} and enforcing them by design is a promising avenue for future work.\looseness=-1

    \section{Conclusion}

In this paper, we showed how to overcome the performance limitations of model-based relational tools for speculative leaks.
This allowed us to drastically speed up black-box testing for speculative leaks and to run a testing campaign of a previously-infeasible depth (130 million test executions) on Intel and AMD CPUs.
Our testing campaign discovered two new leaks that have escaped previous analyses, which demonstrates the potential for automated detection of speculative leaks.

\section*{Acknowledgment}

We would like to thank Saar Amar, Amaury Chamayou, Manuel Costa, C{\'e}dric Fournet, Istvan Haller, and Stavros Volos for discussions and support. We'd like to thank the anonymous reviewers and Scott. D. Constable for feedback on the paper.
This work was partially supported 
by the Madrid regional government under the project S2018/TCS-4339 BLOQUES-CM, 
by the Spanish Ministry of Science, Innovation, and University under the projects RTI2018-102043-B-I00 SCUM and TED2021-132464B-I00 PRODIGY and under the Ram\'on y Cajal grant RYC2021-032614-I,  
and by a gift from Intel Corporation.

    \bibliographystyle{IEEEtran.bst}
    \bibliography{ms}
    
    \appendix
    \subsection{Subsets of x86-64 Used in Evaluation}
\label{app:subsets}

In the main testing campaign (\secref{findings}) and in the evaluation (\secref{eval}), most of the experiments were performed on subsets of the x86-64 ISA. 
These particular subsets were selected because together they form a complete base x86-64 instruction set, excluding those instructions that are not (yet) supported by Revizor.
Specifically, we excluded system instructions (e.g., \code{SYSCALL}), instructions incorrectly emulated by Unicorn (e.g., \code{ROR}), and control-flow instructions for which no contracts are available (\code{RET, CALL}, and indirect jumps).
We omitted ISA extensions like AVX or x87 because they require more complex test case generation algorithms to avoid faults, also not yet supported by Revizor.

Each of the tested subsets constituted of several basic arithmetic instructions (including their versions with memory operands) and of several instructions that are unique to the given subset. The exact instructions are as follows:

\begin{itemize}
    \item \code{cond} (conditional branches): \code{\textit{ADC, ADD, CMP, DEC, INC, NEG, SBB, SUB}, J*, LOOP*, JMP} (unconditional direct jump).
    \item \code{strn} (string operations): \code{\textit{ADC, ADD, CMP, DEC, INC, NEG, SBB, SUB}, CLC, CLD, CMC, LAHF, LOCK, REPE, REPNE, SAHF, SCASB, SCASD, SCASW, STC, STD}.
    \item \code{dmul} (division and multiplication): \code{\textit{ADC, ADD, CMP, DEC, INC, NEG, SBB, SUB}, DIV, IMUL, MUL}.
    \item \code{flag} (operations on flags): \code{\textit{ADC, ADD, CMP, DEC, INC, NEG, SBB, SUB}, CLC, CLD, CMC, LAHF, SAHF, STC, STD}.
    \item \code{lock} (atomics with LOCK prefix): \code{\textit{ADC, ADD, CMP, DEC, INC, NEG, SBB, SUB}, LOCK ADC, LOCK ADD, LOCK CMP, LOCK DEC, LOCK INC, LOCK NEG, LOCK SBB, LOCK SUB, LOCK BSF, LOCK BSR, LOCK BT, LOCK BTC, LOCK BTR, LOCK BTS, LOCK NOT, LOCK OR, LOCK TEST, LOCK XOR}.
    \item \code{atom} (atomics without LOCK prefix): \code{\textit{ADC, ADD, CMP, DEC, INC, NEG, SBB, SUB}, CMPXCHG, XADD, LOCK CMPXCHG, LOCK XADD}.
    \item \code{dxfr} (data transfer): \code{\textit{ADC, ADD, CMP, DEC, INC, NEG, SBB, SUB}, BSWAP, MOV, MOVSX, MOVZX, XCHG}.
    \item \code{setc} (conditional byte set): \code{\textit{ADC, ADD, CMP, DEC, INC, NEG, SBB, SUB}, SET*}.
    \item \code{nop} (NOP instructions): \code{\textit{ADC, ADD, CMP, DEC, INC, NEG, SBB, SUB}, NOP}.
    \item \code{logi} (logical operations): \code{\textit{ADC, ADD, CMP, DEC, INC, NEG, SBB, SUB}, AND, NOT, OR, TEST, XOR}.
    \item \code{conv} (data type conversion): \code{\textit{ADC, ADD, CMP, DEC, INC, NEG, SBB, SUB}, CBW, CDQ, CWD, CWDE}.
    \item \code{cmov} (conditional moves): \code{\textit{ADC, ADD, CMP, DEC, INC, NEG, SBB, SUB}, CMOV*}.
    \item \code{bit} (bit test and bit scan): \code{\textit{ADC, ADD, CMP, DEC, INC, NEG, SBB, SUB}, BSF, BSR, BT, BTC, BTR, BTS}.
\end{itemize}

\subsection{Example: A walk through the discovery of SCO}
\label{app:sco}

In this example, we demonstrate the process of discovering a speculative leak with our tool by showing how we found SCO.

As a part of the main fuzzing campaign (\tabref{results1}), we tested the \code{strn} subset of x86-64 on Intel Xeon E-2288G against the \code{CT-SEQ} contract.
This contract permits side-channel information leakage through caches, but only for non-speculatively executed instructions.
Accordingly, if Revizor finds an instance of \emph{any} speculative leakage while testing against this contract, Revizor will report it as a violation.

To set this experiment up, we wrote the following configuration file:

\begin{lstlisting}[aboveskip=5pt, belowskip=5pt]
instruction_categories:
  - BASE-BINARY
  - BASE-STRINGOP
contract_observation_clause: ct
contract_execution_clause:
  - seq
enable_speculation_filter: true
enable_observation_filter: true
inputs_per_class: 2
\end{lstlisting}

Here, lines 1--3 select the \code{strn} subset; lines 4--6 tell Revizor to test against \code{CT-SEQ} contract;
lines 7 and 8 enable the speculation and the observation filters (\subsecref{des-spec} and~\subsecref{des-observ});
and line 9 tells CIG (\subsecref{des-boost}) to create two inputs for each input class.

We passed this configuration file to our modified version of Revizor, and launched a testing campaign with 100'000 programs, each tested with 100 inputs:

\begin{lstlisting}[numbers=none, aboveskip=5pt, belowskip=5pt]
./cli.py fuzz -c conf.yaml -n 100000 -i 100
\end{lstlisting}

Already after starting the testing process, we saw signs of speculation: An unusually high number of test cases passed through the filters.
(\figref{gains-illustration} illustrates this observation.)

After about three hours of testing, Revizor detected a violation.
At this point, we knew that we have discovered a new speculation type---there has been previously no speculative leak reported within this subset of instructions.
However, we did not know the source of leakage.

From Revizor, we received a test case---a program and a sequence of inputs---that witnessed an unexpected information leakage.
As the program was generated randomly, it contained many irrelevant instructions, which complicated the investigation.
Therefore, we passed the program to Revizor's automatic minimizer, and it returned the following:

\begin{lstlisting}[aboveskip=5pt, belowskip=5pt]
CLD  # instrumentation
SUB CL, DL
AND RSI, 0b1111111111111 # instrumentation
SUB RAX, -1904627778
NEG AL
CMP EDX, -122
AND RBX, 0b1111111111111 # instrumentation
SBB byte ptr [R14 + RBX], 111
CMP SI, 117
AND RDI, 0b1111111111111 # instrumentation
ADD RDI, R14 # instrumentation
AND RCX, 0xff # instrumentation
ADD RCX, 1 # instrumentation
REPNE SCASD
\end{lstlisting}

Next, we tried to find out which specific instruction within this program triggers speculation.
For this, we removed one instruction at a time, executed the test case on Revizor, and checked if it still passed the speculation filter.
The culprit turned out to be \code{REPNE SCASD} at line~14.

From there on, we manually reverse-engineered the mechanism behind the speculation of \code{SCAS}, and it led to the discovery of SCO.

\subsection{Configurations Used in Evaluation}
\label{app:configs}

The following is a detailed description of the testing configurations used in~\subsecref{eval-speed} and~\figref{detection}. The names of the ISA subsets are described in~\appref{subsets}.

\myparagraphnodot{Configuration Spectre V1:}
\begin{itemize}
    \item ISA subsets: \code{cond}, \code{nop}, \code{bit}, \code{cmov}, \code{conv}, \code{dxfr}, \code{flag}, \code{setc}, \code{logi}
    \item Target contract: \ctseq{}
    \item Spectre V4 patch enabled: True
    \item Microcode assists permitted: False
\end{itemize}

The configuration can surface V1 because the instruction set contains conditional branches. In principle, this configuration can also detect V1-Var, yet the probability of detecting V1-Var is much lower compared to V1, thus it has only a minor impact on the measurement results.

The configuration cannot surface V4 because the corresponding patch is enabled; LVI---because microcode assists are not permitted; ZDI---because the subsets do not include divisions; SCO---because the subsets do not include string operations.

\myparagraphnodot{Configuration Spectre V4:}
\begin{itemize}
    \item ISA subsets: \code{nop}, \code{bit}, \code{cmov}, \code{conv}, \code{dxfr}, \code{flag}, \code{setc}, \code{logi}
    \item Target contract: \ctseq{}
    \item Spectre V4 patch enabled: False
    \item Microcode assists permitted: False
\end{itemize}

The configuration can surface V4 because the instruction set contains memory accesses and the patch is disabled. The configuration cannot surface V1 and V1-Var because the instruction set does not contain branches; LVI---because microcode assists are not permitted; ZDI---because subsets do not include divisions; SCO---because subsets do not include string operations.

\myparagraphnodot{Configuration LVI-Null:}
\begin{itemize}
    \item ISA subsets: \code{nop}, \code{bit}, \code{cmov}, \code{conv}, \code{dxfr}, \code{flag}, \code{setc}, \code{logi}
    \item Target contract: \ctseq{}
    \item Spectre V4 patch enabled: True
    \item Microcode assists permitted: True
\end{itemize}

The configuration can surface LVI because microcode assists are permitted. The configuration cannot surface V1 and V1-Var because the instruction set does not contain branches; V4---because the corresponding patch is enabled; ZDI---because subsets do not include divisions; SCO---because subsets do not include string operations.

\myparagraphnodot{Configuration V1-Var:}
\begin{itemize}
    \item ISA subsets: \code{nop}, \code{bit}, \code{cond}, \code{cmov}, \code{conv}, \code{dxfr}, \code{flag}, \code{setc}, \code{logi}
    \item Target contract: \ctcond{}
    \item Spectre V4 patch enabled: True
    \item Microcode assists permitted: False
\end{itemize}

The configuration can surface V1-Var because the instruction set contains conditional branches and variable-latency instructions. The configuration cannot surface V4 because the corresponding patch is enabled; LVI---because microcode assists are not permitted; V1 is not reported because the target contract (\ctcond{}) permits conditional branch misprediction; ZDI---because subsets do not include divisions; SCO---because subsets do not include string operations.

\myparagraphnodot{Configuration ZDI:}
\begin{itemize}
    \item ISA subsets: \code{dmul}, \code{nop}, \code{bit}, \code{cond}, \code{cmov}, \code{conv}, \code{dxfr}, \code{flag}, \code{setc}, \code{logi}
    \item Target contract: \ctseq{}
    \item Spectre V4 patch enabled: True
    \item Microcode assists permitted: False
\end{itemize}

The configuration can surface ZDI because the instruction set contains 64-bit divisions. The configuration cannot surface V1 and V1-Var because the instruction set does not contain branches; V4---because the corresponding patch is enabled; LVI---because microcode assists are not permitted; SCO---because subsets do not include string operations.

\myparagraphnodot{Configuration SCO:}
\begin{itemize}
    \item ISA subsets: \code{strn}, \code{nop}, \code{bit}, \code{cond}, \code{cmov}, \code{conv}, \code{dxfr}, \code{flag}, \code{setc}, \code{logi}
    \item Target contract: \ctseq{}
    \item Spectre V4 patch enabled: True
    \item Microcode assists permitted: False
\end{itemize}

The configuration can surface SCO because the instruction set contains string operations. The configuration cannot surface V1 and V1-Var because the instruction set does not contain branches; V4---because the corresponding patch is enabled; LVI---because microcode assists are not permitted; ZDI---because subsets do not include divisions.

\end{document}